\documentclass[prd,showpacs,preprintnumbers,twocolumn,amsmath,nofootinbib,amssymb]{revtex4}
\usepackage{graphicx,color,dcolumn,booktabs,bm}
\usepackage{longtable,lscape}
\usepackage{txfonts}
\usepackage{overpic}
\usepackage{amssymb}
\usepackage{epstopdf}
\usepackage{indentfirst}
\usepackage{feynmf}   %{feynmp}
\usepackage{slashed}  %for Feynman symbols
\usepackage{cases}
\usepackage{color}
\usepackage{float}
\usepackage{multirow}
\usepackage{graphicx,color,dcolumn,booktabs,bm}
\usepackage{epsfig,dsfont,amssymb,amsmath,amsfonts,amsbsy,mathrsfs}
\graphicspath{{Figures/}} %

\usepackage{hyperref}
\hypersetup{colorlinks,citecolor=blue,anchorcolor=red,menucolor=red, linkcolor=red,filecolor=red,runcolor=red,urlcolor=blue,frenchlinks=red}

\makeatletter
\@addtoreset{equation}{section}
\makeatother

\allowdisplaybreaks

\begin{document}
%\begin{CJK}{GBK}{}

\title{$\bar{K}\Lambda$ molecular explanation to the newly observed $\Xi(1620)^0$}

\author{Kan Chen$^{1,2}$}
\email{chenk$_10$@lzu.edu.cn}
\author{Rui Chen$^{1,2,3,4}$}
\email{chenr15@lzu.edu.cn}
\author{Zhi-Feng Sun$^{1,2}$}
\email{sunzf@lzu.edu.cn}
\author{Xiang Liu$^{1,2}$}
\email{xiangliu@lzu.edu.cn}
\affiliation{ $^1$School of Physical
Science and Technology, Lanzhou University, Lanzhou 730000, China
\\
$^2$Research Center for Hadron and CSR Physics, Lanzhou University
and Institute of Modern Physics of CAS, Lanzhou 730000, China\\
$^3$School of Physics and State Key Laboratory of Nuclear Physics and Technology, Peking University, Beijing 100871, China\\
$^4$Center of High Energy Physics, Peking University, Beijing 100871, China}

\begin{abstract}
  The newly observed $\Xi(1620)^0$ by the Belle Collaboration inspires our interest in performing a systematic study on the interaction of an antistrange meson $(\bar{K}^{(*)})$ with a strange or doubly strange ground octet baryon $\mathcal{B}$ ($\Lambda$, $\Sigma$, and $\Xi$), where the spin-orbit force and the recoil correction are considered in the adopted one-boson-exchange model. Our results indicate that $\Xi(1620)^0$ can be explained as a $\bar{K}\Lambda$ molecular state with $I(J^P)=1/2(1/2^-)$ and the intermediate force from $\sigma$ exchange plays an important role. Additionally, we also predict several other possible molecular candidates, i.e., the $\bar{K}\Sigma$ molecular state with $I(J^P)=1/2(1/2^-)$ and the triply strange $\bar{K}\Xi$ molecular state with $I(J^P)=0(1/2^-)$.
\end{abstract}

\pacs{14.20.Pt, 12.39.Pn, 13.75.Jz}

\maketitle

\section{introduction}\label{sec1}

Recently, the Belle Collaboration \cite{Sumihama:2018moz} announced the observation of $\Xi(1620)^0$ in the $\Xi_c^-\pi^+$ invariant mass spectrum of the $\Xi_c^+\to \Xi^-\pi^+\pi^+$ process, which confirmed early experimental evidence of $\Xi(1620)$ existing in the $K^-p$ reaction in the 1970s \cite{Briefel:1977bp,Bellefon:1900zz,Ross:1972bf}.
The measured mass and width are
\begin{eqnarray*}
M &=& 1610.4\pm 6.0~\text{(stat)}_{-3.5}^{+5.9}~\text{(syst)}~\text{MeV},\\
\Gamma &=& 59.9\pm4.8~\text{(stat)}_{-3.0}^{+2.8}~\text{(syst)}~\text{MeV},
\end{eqnarray*}
respectively. Besides observing the $\Xi(1620)^0$ signal, Belle also first reported the evidence of $\Xi(1690)^0$ in the $\Xi_c^+\to \Xi^-\pi^+\pi^+$ process  \cite{Sumihama:2018moz}.

\begin{figure}[!htbp]
\center
\includegraphics[width=3.5in]{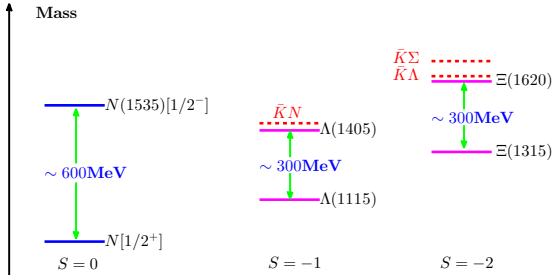}\\
\caption{A comparison between the masses of the molecular candidates and the mass thresholds of a pair of antimeson and baryon systems.}\label{mass}
\end{figure}

Focusing on $\Xi(1620)^0$, we must mention the similarities between $\Xi(1620)^0$ and the famous $\Lambda(1405)$. As shown in Fig. \ref{mass}, we list the mass gaps of several typical states and the corresponding thresholds. We notice that the mass gap between $\Xi(1620)$
 and $\Xi(1315)$ is similar to that of $\Lambda(1405)$
 and $\Lambda(1115)$, where $\Lambda(1115)$ and $\Xi(1315)$ are ground states with $J^P=1/2^+$ in the corresponding $\Lambda$ and $\Xi$ baryon families. And, these two mass gaps around 300 MeV are two times smaller than the mass gap between the $N(1535)$ and the nucleon.
These phenomena show that $\Lambda(1405)$ and $\Xi(1620)^0$ are not consistent with the predicted
masses of $\Lambda(1/2^-)$ and $\Xi(1/2^-)$ from quark model \cite{Capstick:1986bm}.
What is more special is that $\Xi(1620)^0$ and $\Lambda(1405)$ are just below the $\bar{K}N$ and the $\bar{K}\Lambda$ thresholds, respectively.

Since M. Gell-Mann \cite{GellMann:1964nj} and G. Zweig \cite{Zweig:1981pd} first proposed the existence of the exotic states in their pioneer work on the quark model, great theoretical and experimental efforts were made on searching for exotic hadronic matter. The studies of exotic hadronic matter can deepen our understanding of the nonperturbative behavior of QCD.
As an important configuration of exotic hadronic matter, hadronic molecules have received extensive attentions in the past decade \cite{Chen:2016qju,Liu:2019zoy,Guo:2017jvc}. In particular, the updated analysis from the LHCb Collaboration indicated the observation of three near threshold hidden-charm pentaquarks, $P_c(4312)$, $P_c(4440)$, and $P_c(4457)$ \cite{Aaij:2019vzc}, which provides a strong evidence for the existence of hidden-charm meson-baryon configuration molecular states \cite{Wu:2010jy,Yang:2011wz,Chen:2019asm,Chen:2019bip,Liu:2019tjn,He:2019ify,Huang:2019jlf,Weng:2019ynv}.

In fact, there is a long-term discussion on the meson-baryon molecules in the light flavor sector. $\Lambda(1405)$ is a typical example which was assigned as a $\bar{K}N$ molecular candidate with $I(J^P)=0(1/2^-)$, since its mass is close to the $\bar{K}N$ threshold but far away from the prediction in quark model (see the review articles \cite{Klempt:2009pi,Klempt:2007cp} for details).

Due to these similarities between $\Xi(1620)$ and $\Lambda(1405)$, it is interesting to study whether the $\Xi(1620)$ can be the doubly strange molecular partner of the $\Lambda(1405)$. In Refs. \cite{Ramos:2002xh,Miyahara:2016yyh}, the $\Xi(1620)$ was interpreted as a $J^P=1/2^-$ resonance dynamically generated in chiral unitary approach. By introducing the vector exchange interaction, the Bethe-Salpeter equation approach \cite{Wang:2019krq} was applied to identify the $\Xi(1620)$ as a $\bar{K}\Lambda$ or a $\bar{K}\Sigma$ molecular state.

In this work, we will discuss the $\Xi(1620)$ as a molecular state in the framework of the one-boson-exchange (OBE) model. In general, the spin-orbit force and the recoil correction are very important for hadron-hadron interactions in the light flavor sector, which will be also included in the following calculations. As a byproduct, the $\bar{K}^{(*)}\mathcal{B}$ $(\mathcal{B}=\Lambda/\Sigma/\Xi)$ systems will also be investigated. %{\color{red}Note that, in the $\bar{K}^*\mathcal{B}$ systems, the $\bar{K}^*$ can decay into $K^-\pi^+$. Taking the same arguments proposed in Ref. \cite{Zhao:2013xha,Chen:2015add}, such molecular resonances still have possibilities to form from the $\bar{K}^*\mathcal{B}$ interactions. Theoretically, for the studied $\bar{K}^{*}\mathcal{B}$ systems, we can approximately treat the $\bar{K}^*$ as point-like particle.} %Here, we check the possibilities for the existence of other possible molecular candidates with strange numbers $S=-2$ and $-3$.

%When study the assignments of light baryon states, an important %way is to distinguish the exotic states from these conventional %baryons. Thus, our work is crucial to clarify these exotic states.

This paper is organized as follows. In Sec. \ref{sec2}, we present the
deduction of the OBE effective potentials. In Sec. \ref{sec3}, the corresponding numerical results and discussion for $\bar{K}^{(*)}\mathcal{B}$ systems are given. This paper will end with a summary in Sec. \ref{sec4}.

\section{Interactions}\label{sec2}

%\subsection{Effective Lagrangians}
In the local hidden gauge approach \cite{Garzon:2012np,Lu:2016nlp}, the effective Lagrangians depicting the interaction of vector mesons, vector meson with pseudoscalar mesons can be constructed as
\begin{eqnarray}
\label{VVV}
\mathcal{L}_{VVV} &=& ig\left\langle\left(\partial_{\mu}V_{\nu}-\partial_{\nu}V_{\mu}\right)
V^{\mu}V^{\nu}\right\rangle,\\
\label{PPV}
\mathcal{L}_{PPV} &=& -ig\left\langle\left[P,\partial_{\nu}P\right]V^{\nu}\right\rangle,\\
\label{VVP}
\mathcal{L}_{VVP} &=& \frac{G}{\sqrt{2}}\epsilon^{\mu\nu\alpha\beta}
\left\langle\partial_{\mu}V_{\nu}\partial_{\alpha}V_{\beta}P\right\rangle.
\end{eqnarray}
In the above Lagrangians, $g=\frac{12}{2\sqrt{2}}$ and $G=\frac{3g^2}{4\pi^2f_{\pi}}$ with $f_{\pi}=93$ MeV were given in Ref. \cite{Lu:2016nlp}.
In Ref. \cite{Garzon:2012np}, the lowest order baryon meson Lagrangians are expressed as
\begin{eqnarray}
\label{BBP}
\mathcal{L}_{BBP}&=&-\frac{D+F}{\sqrt{2}f_{\pi}}\left\langle\bar{B}\gamma_{\mu}\gamma_5
\partial^{\mu}PB\right\rangle-\frac{D-F}{\sqrt{2}f_{\pi}}
\left\langle\bar{B}\gamma_{\mu}\gamma_5B\partial^{\mu}P\right\rangle,\nonumber\\
\mathcal{L}_{BBV} &=& g\left(\left\langle\bar{B}\gamma_{\mu}\left[V^{\mu},B\right]\right\rangle
+\left\langle\bar{B}\gamma_{\mu}B\right\rangle\left\langle V^{\mu}\right\rangle\right),
\end{eqnarray}
where $D=0.75$ and $F=0.51$ \cite{Garzon:2012np}. Here, matrixes for vector mesons, pseudoscalar mesons, and light baryons in SU(3) octet are respectively written as
\begin{eqnarray}
{V} &=& \left(\begin{array}{ccc}
\frac{\rho^0}{\sqrt{2}}+\frac{\omega}{\sqrt{2}}  &\rho^+      &K^{*+}\\
\rho^- &-\frac{\rho^0}{\sqrt{2}}+\frac{\omega}{\sqrt{2}}      &K^{*0}\\
K^{*-}      &\bar{K}^{*0}    &\phi
\end{array}\right),\nonumber\\
{P} &=& {\left(\begin{array}{ccc}
       \frac{\pi^0}{\sqrt{2}}+\frac{\eta}{\sqrt{6}} &\pi^+      &K^+\\
       \pi^-       &-\frac{\pi^0}{\sqrt{2}}+\frac{\eta}{\sqrt{6}}     &K^0\\
       K^-         &\bar{K}^0      &-\frac{2}{\sqrt{6}}\eta
               \end{array}\right)},\nonumber\\
B &=& \left(\begin{array}{ccc}
\frac{\Sigma^0}{\sqrt{2}}+\frac{\Lambda}{\sqrt{6}}    &\Sigma^+    &p\\
\Sigma^-  &-\frac{\Sigma^0}{\sqrt{2}}+\frac{\Lambda}{\sqrt{6}}   &n\\
\Xi^-      &\Xi^0     &-\frac{2}{\sqrt{6}}\Lambda\end{array}\right).\nonumber
\end{eqnarray}

In the framework of the OBE model, the intermediate-range interaction is provided by the $\sigma$ exchange process. This scalar meson exchange contribution has been widely introduced to study various other multiquark molecular bound states \cite{Chen:2019asm,Yang:2011wz,Ding:2004ty,Wang:2019aoc,Chen:2017jjn,Chen:2018pzd,Chen:2017xat,Chen:2016ryt,Liu:2011xc,Meng:2017fwb}. The corresponding Lagrangians are
\begin{eqnarray}
\label{PPS}
\mathcal{L}_{PP\sigma} &=& g_{\sigma }m_P\left\langle{P}P\sigma\right\rangle,\\
\mathcal{L}_{VV\sigma} &=& g_{\sigma }m_V\left\langle{V}V\sigma\right\rangle,\\
\label{VVS}
\mathcal{L}_{BB\sigma} &=& g_{\sigma }^{\prime}\left\langle
\bar{B}\sigma B\right\rangle.
\label{BBS}
\end{eqnarray}
Here, $m_{V}$ and $m_P$ denote the masses of vector and pseudoscalar mesons, respectively. In quark model, the coupling constants in Eqs. (\ref{PPS})-(\ref{BBS}) have the relation of $g^\prime_{\sigma}=g_{\sigma NN}=\frac{3}{2}g_{\sigma}$. In Ref. \cite{Wang:2019aoc}, $\frac{g_{\sigma NN}^2}{4\pi}=5.69$ was determined.

\begin{figure}[!htbp]
\center
\includegraphics[width=3.0in]{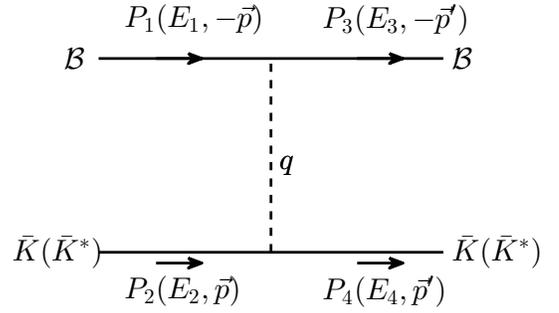}\\
\caption{Feynman diagram for the $\bar{K}^{(*)}\mathcal{B}\rightarrow \bar{K}^{(*)}\mathcal{B}$ process.}\label{feynman}
\end{figure}

%\subsection{Effective Potential}\label{subpotential}
With the Lagrangians given in Eqs. (\ref{VVV})-(\ref{BBS}), we can derive the scattering amplitude for the $\bar{K}^{(*)}\mathcal{B}\to\bar{K}^{(*)}\mathcal{B}$ process in $t-$channel. In Fig. \ref{feynman}, we present the corresponding Feynman diagram and the four momentum for the initial and the final states. The OBE effective potential $\mathcal{V}_E^{h_1h_2\rightarrow h_3h_4}(\vec{q})$ can be related to the scattering amplitude for the process $h_1h_2\rightarrow h_3h_4$ via the Breit approximation, i.e.,
\begin{eqnarray}
\mathcal{V}_{E}^{h_1h_2\rightarrow h_3h_4}(\vec{q})=-\frac{\mathcal{M}\left(h_1h_2\rightarrow h_3h_4\right)}{\sqrt{\prod_i2M_i\prod{f}2M_f}},
\end{eqnarray}
where $\mathcal{M}\left(h_1h_2\rightarrow h_3h_4\right)$ is the scattering amplitude. $M_i$ and $M_f$ denote the masses of initial and final states, respectively.

And then, we get the OBE effective potentials for the $\bar{K}\mathcal{B}^{(*)}$ systems
\begin{eqnarray}
V_{\sigma}&=&\frac{-g_{\sigma} g^\prime_{\sigma}}{\vec{q}^2+m^2_{\sigma}}\left[1-\frac{\vec{k}^2}{2m^2_{\mathcal{B}}}
-\frac{1}{4m^2_{\mathcal{B}}}i\vec{\sigma}\cdot\left(\vec{q}\times\vec{k}\right)\right],\\
V_{V}&=&-\frac{g^2}{\vec{q}^2+m_V^2}\left[1-\frac{\vec{q}^2}{8m_{\mathcal{B}}^2}+\frac{\vec{k}^2}
{m_{K}m_{\mathcal{B}}}\right.\nonumber\\
&&\left.+\frac{m_K+2m_\mathcal{B}}{4m_{\mathcal{B}}^2m_K}i\vec{\sigma}\cdot
\left(\vec{q}\times\vec{k}\right)\right],\\
V^\prime_{\sigma}&=&-\frac{g_{\sigma}g^\prime_{\sigma}}{\vec{q}^2
+m_{\sigma}^2}\left(\vec{\epsilon}_4^{\dagger\prime}
\cdot\vec{\epsilon}_2\right)\left[1-\frac{\vec{k}^2}{2m_\mathcal{B}^2}-\frac{i\vec{\sigma}
\cdot\left(\vec{q}\times\vec{k}\right)}{4m_\mathcal{B}^2}\right],\\
V^\prime_V&=&-\frac{g^2}{\vec{q}^2+m_{V}^2}\Bigg\{\left(\vec{\epsilon}_4^{\dagger\prime}
\cdot\vec{\epsilon}_2\right)\left[1-\frac{\vec{q}^2}{m_\mathcal{B}^2}+\frac{\vec{k}^2}
{m_{K^*}m_{\mathcal{B}}}\right]
\nonumber\\&&-\frac{\left(\vec{\epsilon}_4^{\dagger\prime}\cdot\vec{q}\right)
\left(\vec{\epsilon}_2\cdot\vec{q}\right)}{2m_{K^*}^{2}}
+\frac{m_{K^*}+2m_\mathcal{B}}{4m_{\mathcal{B}}^2m_{K^*}}\left(\vec{\epsilon}_4^{\dagger\prime}
\cdot\vec{\epsilon}_2\right)\left[i\vec{\sigma}\cdot\left(\vec{q}\times\vec{k}\right)
\right]\nonumber\\
&&+\frac{i\left[\left(\vec{\epsilon}_4^{\dagger\prime}\times\vec{\epsilon}_2\right)
\times\vec{q}\right]
\cdot\left(\vec{\sigma}\times\vec{q}\right)}{2m_{K^*}m_{\mathcal{B}}}
\nonumber\\&&+\frac{m_{\mathcal{B}}+2m_{K^*}}{2m_{\mathcal{B}}m_{K^*}^2}
\left(\vec{\epsilon}_4^{\dagger\prime}\times\vec{\epsilon}_2\right)\cdot
\left(\vec{q}\times\vec{k}\right)\Bigg\},\\
V_P&=&-\frac{\sqrt{2}G}{4\left(\vec{q}^2+m^2_{P}\right)}
i\left[\left(\vec{\epsilon}_4^{\dagger\prime}\times\vec{\epsilon}_2\right)\cdot\vec{q}\right]
\left(\vec{\sigma}\cdot\vec{q}\right).
\end{eqnarray}
Here, $V_{\sigma}$ and $V_{V}$ are $\sigma$ exchange and vector exchange potentials for the $\bar{K}\mathcal{B}\rightarrow\bar{K}\mathcal{B}$ processes, respectively. While in the $\bar{K}^*\mathcal{B}\rightarrow \bar{K}^*\mathcal{B}$ processes, $\sigma$ exchange, vector exchange, and pseudoscalar exchange potentials are respectively denoted by $V^\prime_\sigma$, $V^\prime_V$, and $V_P$. $\vec{q}=\vec{p}^\prime-\vec{p}$ and $\vec{k}=1/2(\vec{p}+\vec{p}^\prime)$. Additionally, $m_\sigma$, $m_P$, and $m_V$ denote the masses of exchanged scalar meson $(\sigma)$, pseudoscalar mesons $(\pi, \eta)$ and vector mesons $(\rho, \omega, \phi)$, respectively. After performing the Fourier transformation, we may extract the effective potentials in the coordinate space, i.e.,
\begin{eqnarray}
\mathcal{V}=\int \frac{d^3\vec{q}}{\left(2\pi^3\right)}e^{i\vec{q}\cdot\vec{r}}\mathcal{V}_{E}^{h_1h_2\rightarrow h_3h_4}\left(\vec{q}\right)\mathcal{F}^2\left(q^2,m^2_{E}\right).
\end{eqnarray}
Here, the form factor $\mathcal{F}(q^2,m_E^2)=(\Lambda^2-m_{E}^2)/(\Lambda^2-q^2)$ is introduced in every interactive vertex, which can reflect the finite size effect of the discussed hadrons and compensate the off-shell effects of the exchanged mesons. $\Lambda$, $m_E$, and $q$ are the cutoff, mass and four momentum of the exchanged mesons, respectively. According to the experience from deuteron \cite{Ding:2004ty}, the cutoff $\Lambda$ is taken around 1.0 GeV, which is often regarded as a typical cutoff value for a loosely bound hadronic molecular state.

Since the $S-D$ wave mixing effect is considered, the spin-orbital wave functions for the $\bar{K}^{(*)}\mathcal{B}$ systems with quantum numbers $J^P$ can be written as
\begin{eqnarray*}
\bar{K}\mathcal{B}\left(\frac{1}{2}^-\right):&& \left|{}^2S_{\frac{1}{2}}\right\rangle,\\
\bar{K}^*\mathcal{B}\left(\frac{1}{2}^-\right): &&\left|{}^2S_{\frac{1}{2}}\right\rangle,\quad \left|{}^4D_{\frac{1}{2}}\right\rangle,\\
\bar{K}^*\mathcal{B}\left(\frac{3}{2}^-\right):
&&\left|{}^4S_{\frac{3}{2}}\right\rangle,\quad \left|{}^4D_{\frac{3}{2}}\right\rangle,\quad \left|{}^4D_{\frac{3}{2}}\right\rangle.
%J^P&=\frac{5}{2}^-:{}^2D_{\frac{5}{2}},\qquad\quad
%J^P&=&\frac{5}{2}^-:{}^2D_{\frac{5}{2}}, {}^4D_{\frac{5}{2}},\nonumber
\end{eqnarray*}
The expansions of the spin-orbital wave functions $|\bar{K}^{(*)}\mathcal{B}\left({}^{2S+1}L_J\right)\rangle$ are
\begin{eqnarray*}
\quad\left|\bar{K}\mathcal{B}\left({}^{2S+1}L_J\right)\right\rangle&=&\sum_{m,m^\prime} C_{Sm_S,Lm_L}^{J,M}\chi_{\frac{1}{2}m}\left|Y_{L,m_L}\right\rangle,\\
\quad\left|\bar{K}^*\mathcal{B}\left({}^{2S+1}L_J\right)\right\rangle&=&
\sum_{\lambda,\lambda^\prime,m_S,m_L} C_{\frac{1}{2}\lambda,1\lambda^\prime}^{S,m_S}C_{Sm_S,Lm_{L}}^{JM}
\chi_{\frac{1}{2}\lambda}\epsilon_{\lambda^\prime}\left|Y_{L,m_{L}}\right\rangle.
\end{eqnarray*}
Here, $C^{J,M}_{S m_S,L m_L}$ and  $C_{\frac{1}{2}\lambda,1\lambda^{\prime}}^{S,m_s}$ are the Clebsch-Gordan coefficients. $\chi_{\frac{1}{2}\lambda}$ and $Y_{L,m_L}$ denote the spin wave function and the spherical harmonic function, respectively. $\epsilon_{\lambda^\prime}$ is the polarization vector of a vector meson in the laboratory frame \cite{Greiner}, with the explicit expression
\begin{eqnarray}
\epsilon_{\lambda^\prime}=\left(\frac{\vec{p}\cdot\vec{\epsilon}_{\lambda^\prime}}{m},
\vec{\epsilon}_{\lambda^\prime}+\frac{\left(\vec{p}\cdot\vec{\epsilon}_{\lambda^\prime}\right)
\vec{p}}{m\left(p_0+m\right)}\right),
\end{eqnarray}
where $p=\left(p_0,\vec{p}\right)$ is the four-momentum in the laboratory frame and $m$ denotes the mass of vector meson.

The detailed Fourier transformations for different types of effective potentials are expressed as \cite{Zhao:2014gqa}
\begin{eqnarray}
\label{Y}
FT\left\{\frac{1}{\vec{q}^2+m^2}\left(\frac{\Lambda^2-m^2}
{\Lambda^2+\vec{q}^2}\right)^2\right\}&=&Y\left(\Lambda,m,r\right),\\
FT\left\{\frac{\vec{q}^2}{\vec{q}^2+m^2}\left(\frac{\Lambda^2-m^2}
{\Lambda^2+\vec{q}^2}\right)^2\right\}&=&-\nabla^2Y\left(\Lambda,m,r\right),\\
\label{Z}
FT\left\{\frac{\vec{k}^2}{\vec{q^2}+m^2}\left(\frac{\Lambda^2-m^2}
{\Lambda^2+\vec{q}^2}\right)^2\right\}\nonumber&=&\frac{1}{4}\nabla^2Y\left(\Lambda,m,r\right)
\\&&-\frac{1}{2}\left\{\nabla^2,Y\left(\Lambda,m,r\right)\right\}.\nonumber\\
\label{K}
\end{eqnarray}
Here, the $\vec{k}^2$ term in Eq .(\ref{K}) is named as the recoil correction term, and the function $Y\left(\Lambda,m,r\right)$ is defined as
\begin{eqnarray}
Y\left(\Lambda,m,r\right)&=&\frac{1}{4\pi r}\left(e^{-mr}-e^{-\Lambda r}\right)-\frac{\Lambda^2-m^2}{8\pi\Lambda}e^{-\Lambda r}.
\end{eqnarray}
In the above effective potentials, we also introduce several spin-spin interaction operators $\mathcal{D}_1$, $\mathcal{D}_2$, spin-orbital operators $\mathcal{E}_1$, $\mathcal{E}_2$, $\mathcal{E}_3$, and tensor operators $\mathcal{F}_{1}$, $\mathcal{F}_2$. The explicit forms of these operators are
\begin{eqnarray*}\left.\begin{array}{ll}
\mathcal{D}_1=\chi_3^\dagger\vec{\epsilon}^\dagger_4\cdot\vec{\epsilon}_2\chi_1,
&\mathcal{D}_2=\chi_3^\dagger i\left(\vec{\epsilon}^\dagger_4\times\vec{\epsilon}_2\right)
\cdot\vec{\sigma}\chi_1,\\
\mathcal{E}_1=\chi^\dagger_3i\left(\vec{\epsilon}^\dagger_4\times\vec{\epsilon}_2\right)
\cdot\vec{L}\chi_1,
&\mathcal{E}_2=\chi^\dagger_3\vec{\sigma}\cdot\vec{L}\chi_1,\\
\mathcal{E}_3=\chi^\dagger_3\left(\vec{\epsilon}^\dagger_4\cdot\vec{\epsilon}_2\right)
\left(\vec{\sigma}\cdot\vec{L}\right)\chi_1,
&\mathcal{F}_1=\chi^\dagger_3S\left(\vec{\hat{r}},\vec{\epsilon}_4^\dagger,\vec{\epsilon}_2\right)
\chi_1,\\
\mathcal{F}_2=\chi^\dagger_3S\left(\vec{\hat{r}},i(\vec{\epsilon}_4^\dagger
\times\vec{\epsilon}_2),\vec{\sigma}\right)\chi_1,\quad\quad
\end{array}\right.\end{eqnarray*}
where $S(\vec{\hat{r}},\vec{x},\vec{y})$ is the tensor force operator
$S(\vec{\hat{r}},\vec{x},\vec{y})=3(\vec{\hat{r}}\cdot\vec{x})
(\vec{\hat{r}}\cdot\vec{y})-\vec{x}\cdot\vec{y}$,
with $\vec{\hat{r}}=\vec{r}/|\vec{r}|$. In Table \ref{operators}, we present the numerical matrices for these operators.

\renewcommand\tabcolsep{0.3cm}
\renewcommand{\arraystretch}{1.8}
\begin{table}[!htbp]
\caption{The matrix elements for all the operators.}
\centering
\label{operators}
\begin{tabular}{cccccccccccc}
\toprule[1.5pt]
$J^P$  &$\langle\mathcal{D}_1\rangle$ &$\langle\mathcal{D}_2\rangle$
&$\langle \mathcal{E}_1\rangle$\\
\hline
$\frac{1}{2}^-$ &$\begin{pmatrix}1&0\\0&1\end{pmatrix}$ &$\begin{pmatrix}2&0\\0&-1\end{pmatrix}$ &$\begin{pmatrix}0&0\\0&3\end{pmatrix}$\\
\hline
$\frac{3}{2}^-$ &$\begin{pmatrix}1&0&0\\0&1&0\\0&0&1\end{pmatrix}$ &$\begin{pmatrix}-1&0&0\\0&2&0\\0&0&-1\end{pmatrix}$
 &$\begin{pmatrix}0&0&0\\0&2&-1\\0&-1&2\end{pmatrix}$\\
\midrule[1.5pt]
$J^P$&$\langle\mathcal{E}_3\rangle$ &$\langle\mathcal{F}_1\rangle$ &$\langle\mathcal{F}_2\rangle$\\
\hline
$\frac{1}{2}^-$
    &$\begin{pmatrix}0&0\\0&-3\end{pmatrix}$
    &$\begin{pmatrix}0&-2\sqrt{2}\\-2\sqrt{2}&2\end{pmatrix}$
    &$\begin{pmatrix}0&\sqrt{2}\\\sqrt{2}&2\end{pmatrix}$\\
\hline
$\frac{3}{2}^-$ &$\begin{pmatrix}0&0&0\\0&1&-2\\0&-2&-2\end{pmatrix}$ &$\begin{pmatrix}0&2&-2\\2&0&-2\\-2&-2&0\end{pmatrix}$ &$\begin{pmatrix}0&-1&-2\\-1&0&1\\-2&1&0\end{pmatrix}$\\
\bottomrule[1.5pt]
\end{tabular}
\end{table}

With the above preparation, we obtain the total effective potentials for the $\bar{K}^{(*)}\mathcal{B}$ systems
\begin{eqnarray}
\newcommand{\myfont}{\fontsize{2.5pt}{\baselineskip}\selectfont}
\label{VKN}
V_{\bar{K}N}^I&=&V_{\sigma}+\mathcal{G}(I)V_{\rho}+\frac{3}{2}V_{\omega},\\
\label{VKL}
V_{\bar{K}\Lambda}&=&V_{\sigma}+V_{\omega}-V_{\phi},\\
V_{\bar{K}^*\Lambda}&=&V^\prime_\sigma+V^\prime_{\omega}-V^\prime_{\phi},\\
V^{I}_{\bar{K}\Sigma}&=&V_{\sigma}+\mathcal{H}(I)V_{\rho}+V_{\omega}-V_{\phi},\\
V^I_{\bar{K}^*\Sigma}&=&V^\prime_{\sigma}+\mathcal{H}(I)V^{\prime}_\rho
+V^\prime_{\omega}-V^\prime_{\phi}+\mathcal{H}(I)\frac{g_1-g_2}{2}V_{\pi}
\nonumber\\&&-\frac{1}{6}(g_1+g_2)V_{\eta},\\
V^I_{\bar{K}\Xi}&=&V_\sigma+\mathcal{G}(I)V_{\rho}+\frac{1}{2}V_{\omega}-2V_{\phi},\\
V^I_{\bar{K}^*\Xi}&=&V^\prime_{\sigma}+\mathcal{G}(I)V^\prime_{\rho}+\frac{1}{2}
V^\prime_{\omega}-2V^\prime_{\phi}+\mathcal{G}(I)
(-g_2)V_{\pi}\nonumber\\&&+\frac{1}{6}(2g_1-g_2)V_{\eta},
\end{eqnarray}
where $g_1=-\frac{D_1+F_1}{\sqrt{2}f_\pi}$ and $g_2=-\frac{D_1-F_1}{\sqrt{2}f_\pi}$. $\mathcal{H}(I)$ and $\mathcal{G}(I)$ are the isospin factors
\begin{eqnarray*}\left.\begin{array}{ll}
\mathcal{H}\left(I=1/2\right)=2,\quad\quad   &\mathcal{H}\left(I=3/2\right)=-1,\\
\mathcal{G}\left(I=0\right)=3/2,   &\mathcal{G}\left(I=1\right)=-1/2.\end{array}\right.
\end{eqnarray*}
The flavor wave functions for the $\bar{K}^{(*)}\mathcal{B}$ systems are collected in Table \ref{flavor}.
\renewcommand\tabcolsep{0.35cm}
\renewcommand{\arraystretch}{1.8}
\begin{table}[!htbp]
\caption{The flavor wave functions for the discussed $\bar{K}^{(*)}\mathcal{B}$ systems, where $\mathcal{B}$ denotes strange or doubly strange ground octet baryons ($\Lambda$, $\Sigma$, and $\Xi$).}
\centering
\label{flavor}
\begin{tabular}{c|p{30pt}|cccccccccc}
\toprule[1.5pt]
Systems &$|I,I_3\rangle$ &Flavor wave functions\\\hline
\multirow{1}{*}{$\bar{K}N$}
&$|0,0\rangle$&$\frac{1}{\sqrt{2}}\left(\bar{K}^{0}n+K^{-}p\right)$\\
\hline
\multirow{2}{*}{$\bar{K}^{(*)}\Lambda$}&$|\frac{1}{2},\frac{1}{2}\rangle$&$\bar{K}^{(*)0}\Lambda$\\
&$|\frac{1}{2},-\frac{1}{2}\rangle$&$K^{(*)-}\Lambda$\\
\hline
\multirow{6}{*}{$\bar{K}^{(*)}\Sigma$}&$|\frac{3}{2},\frac{3}{2}\rangle$&$\bar{K}^{(*)0}\Sigma^+$\\
&$|\frac{3}{2},\frac{1}{2}\rangle$&$\sqrt{\frac{1}{3}}K^{(*)-}\Sigma^+
-\sqrt{\frac{2}{3}}\bar{K}^{(*)0}\Sigma^0$&\\
&$|\frac{3}{2},-\frac{1}{2}\rangle$&$\sqrt{\frac{2}{3}}K^{(*)-}\Sigma^0
-\sqrt{\frac{1}{3}}\bar{K}^{(*)0}\Sigma^-$&\\
&$|\frac{3}{2},-\frac{3}{2}\rangle$&$K^{(*)-}\Sigma^-$\\
&$|\frac{1}{2},\frac{1}{2}\rangle$&$\sqrt{\frac{2}{3}}K^{(*)-}\Sigma^+
+\sqrt{\frac{1}{3}}\bar{K}^{(*)0}\Sigma^0$\\
&$|\frac{1}{2},-\frac{1}{2}\rangle$&$\sqrt{\frac{2}{3}}\bar{K}^{(*)0}\Sigma^-
+\sqrt{\frac{1}{3}}K^-\Sigma^0$\\
\hline
\multirow{4}{*}{$\bar{K}^{(*)}\Xi$}&$|1,1\rangle$&$\bar{K}^{(*)0}\Xi^0$\\
&$|1,0\rangle$&$\frac{1}{\sqrt{2}}\left(K^{(*)-}\Xi^0-\bar{K}^{(*)0}\Xi^-\right)$\\
&$|1,-1\rangle$&$K^{(*)-}\Xi^-$\\
&$|0,0\rangle$&$\frac{1}{\sqrt{2}}\left(\bar{K}^{(*)0}\Xi^-+K^{(*)-}\Xi^0\right)$\\
\bottomrule[1.5pt]
\end{tabular}
\end{table}

\section{Numerical Results}\label{sec3}
After obtaining the effective potentials and solving the Schr\"{o}dinger equations, we first study whether the newly observed $\Xi(1620)$ can be assigned as a $\bar{K}\Lambda$ molecular state with $I(J^P)=1/2(1/2^-)$. In addition, other possible doubly strange and triply strange $\bar{K}^{(*)}\mathcal{B}$ molecular candidates will be predicted.
%After we have these effective potentials in Sec. \ref{sec2},  we firstly check if the newly $\Xi(1620)$ can be possible doubly strange molecular state composed by $\bar{K}\Lambda$ state with $I(J^P)=1/2(1/2^-)$ by solving the Schr\"{o}dinger equation. In addition, other possible doubly strange and triply strange $\bar{K}^{(*)}\mathcal{B}$ molecular candidates will be predicted in Sec. \ref{KSB}.

\subsection{$\bar{K}\mathcal{B}$ molecules and the $\Xi(1620)$ }\label{KNKL}
\begin{figure}[htbp]
\center
\includegraphics[width=3.4in]{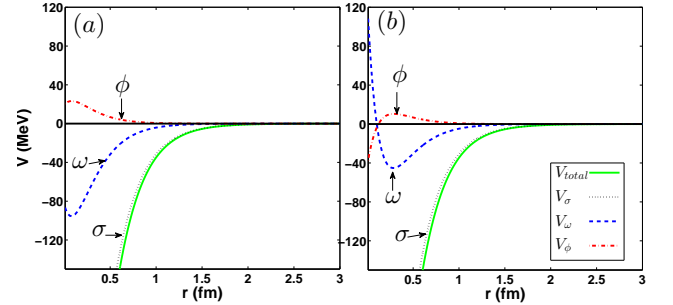}\\
\caption{The $r$ dependence of the OBE effective potentials for the $\bar{K}\Lambda$ system with quantum number $I(J^P)=1/2(1/2^-)$. Diagrams (a) and (b) present the OBE effective potentials without and with the recoil correction effect, respectively.}
\label{potential}
\end{figure}

For the $\bar{K}\Lambda$ system, there does not exist the $\pi/\eta/\rho$ exchange process due to the spin-parity conservation. As shown in Fig. \ref{potential} (a), we present the OBE potentials for the $\Bar{K}\Lambda$ system with $I(J^P)=1/2(1/2^-)$ which depends on $r$. We need to emphasize that we ignore the contribution from the recoil correction. We can see that the dominant $\sigma$ exchange and $\omega$ exchange interactions are both attractive, while the $\phi$ exchange is weakly repulsive.

From Fig. \ref{potential} (b) we can see that the recoil correction only has obvious contribution in the short distance. Since the second term proportional to $\{\nabla^2, Y(\Lambda, m, r)\}$ in Eq. (\ref{K}) should be calculated by acting on the wave functions,
we plot the potentials neglecting the corresponding terms in Fig. \ref{potential} (b). Comparing Fig. \ref{potential} (b) with Fig. \ref{potential} (a), we may see that the recoil correction significantly changes the line shape of $\phi$ and $\omega$ exchange potentials at $r\leq0.5$ fm.
%For the $\bar{K}\Lambda$ system, there is no $\pi/\eta/\rho$ exchange process due to spin-parity conservation. As shown in Fig. \ref{potential} (a), we present the $r$ dependence of OBE effective potentials for $\bar{K}\Lambda [I(J^P)=1/2(1/2)^-]$ system without considered the recoil correction, then we see that the $\sigma$ exchange interaction is attractive and dominant, the $\omega$ exchange interaction is attractive, whereas the $\phi$ exchange is weak repulsive. These results are obviously consist with the conclusions in our previous paper.

%For the right side in Fig. \ref{potential}, the recoil correction is considered in the OBE effective potential, where the recoil correction effect is arisen from $\vec{k}^2/m^2$ term. Finally, we find that the recoil correction effect only contribute to the effective potentials in the short distance range. For example, the recoil correction provides a repulsive force for $\omega$-exchange potential and an attractive force appears in $\phi$-exchange potential.

\begin{figure}[!htbp]
\center
\includegraphics[width=3.5in]{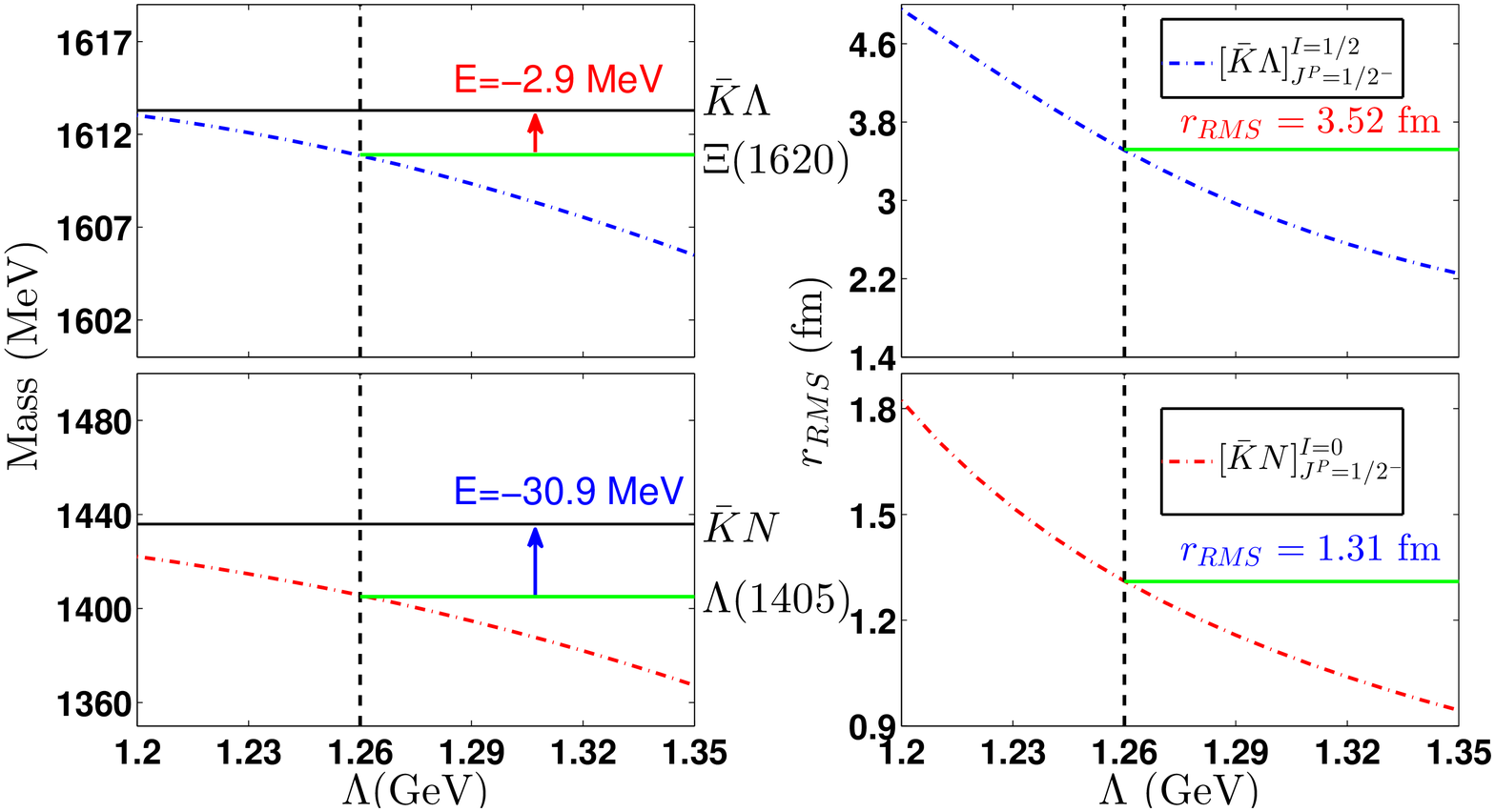}\\
\caption{The $\Lambda$ dependence of the bound state masses and the root-mean-square ($RMS$) radius $r_{RMS}$ for the $\bar{K}N$ and the $\bar{K}\Lambda$ systems with quantum numbers $I(J^P)=0(1/2^-)$ and $I(J^P)=1/2(1/2^-)$, respectively. $E$ is the binding energy. The vertical dotted line and the horizontal solid lines correspond to the cutoff $\Lambda$ value and the thresholds of meson-baryon systems, respectively.}\label{threshold}
\end{figure}

%Before discussing the possibility of the molecular assignment of the newly $\Xi(1620)$, we first reproduce the bound state property of $\Lambda(1405)$, which is composed by the $\bar{K}N$ state with $I(J^P)=0(1/2^-)$. As is well known, cutoff $\Lambda$ corresponds to the typical hadronic scale or the intrinsic size of hadrons, to some extent, one can roughly assume that the cutoff values are the same for the $\bar{K}\mathcal{B}$ $(\mathcal{B}=N, \Lambda, \Sigma, \Xi)$ systems.
As shown in Fig. \ref{threshold}, when the cutoff is taken as $1.26$ GeV, we obtain a $\bar{K}N[0(1/2^-)]$ molecular state with the binding energy $E=-30.9$ MeV and the root-mean-square radius $r_{rms}=1.31$ fm. This molecular state can correspond to the observed $\Lambda(1405)$.

With the same cutoff, we predict that the $\bar{K}\Lambda$ system with $I(J^P)=1/2(1/2^-)$ has the binding energy $-2.9$ MeV, corresponding to the $\Xi(1620)$ observed by the Belle Collaboration \cite{Aaij:2019vzc}. Besides, the binding energy of the $\Lambda(1405)$ is much deeper than that of the $\Xi(1620)$, which can be understood from the obtained potentials. First, from Eq. (\ref{VKN})-(\ref{VKL}), we can know that the $\sigma$ exchange process provides comparable attractive contributions for both $\bar{K}N[0(1/2^-)]$ and $\bar{K}\Lambda[1/2(1/2)^-]$ systems. Besides, for the $\bar{K}N[0(1/2)^-]$ system, $\rho$ and $\omega$ exchange provide attractive forces. However, for the $\bar{K}\Lambda[1/2(1/2)^-]$ system, the allowed exchanged vector mesons include $\omega$ and $\phi$, which provide an attractive and a very weakly repulsive force, respectively. Thus, the interaction of the $\bar{K}N[0(1/2)^-]$ system must be more attractive than that of the $\bar{K}\Lambda[1/2(1/2)^-]$ system.

%If $\Lambda(1405)$ and $\Xi(1620)$ can be possible $\bar{K}N[0(1/2^-)]$ and $\bar{K}\Lambda[1/2(1/2^-)]$ molecular state, respectively. In comparison to their masses and the corresponding meson-baryon threshold, $\Lambda(1405)$ bind much deeper than that in $\Xi(1620)$, their binding energies are around several tens MeV and several MeV, respectively.
%This property can be well understood from their total OBE effective potentials in Eqs. (\ref{VKN}) and (\ref{VKL}). Firstly, the $\sigma$-exchange provides comparable attractive contribution for $\bar{K}N$ and $\bar{K}\Lambda$ systems with $J^P=1/2^-$. And then, for the $\bar{K}N$ system with quantum number $I(J^P)=0(1/2^-)$, $\rho$ and $\omega$ exchange provide attractive force, while for $\bar{K}\Lambda$ system with $1/2(1/2^-)$, the allowed exchanged vector mesons include $\omega$ and $\phi$. Here, although the interaction from $\omega$-exchange is attractive, the $\phi$-exchange plays a very small repulsive interaction role. Thus, the total attractive interaction in $\bar{K}N$ system with $0(1/2^-)$ should be much stronger attractive than that in the $\bar{K}\Lambda$ system with $1/2(1/2^-)$.

\begin{figure}[!htbp]
\center
\includegraphics[width=3.5in]{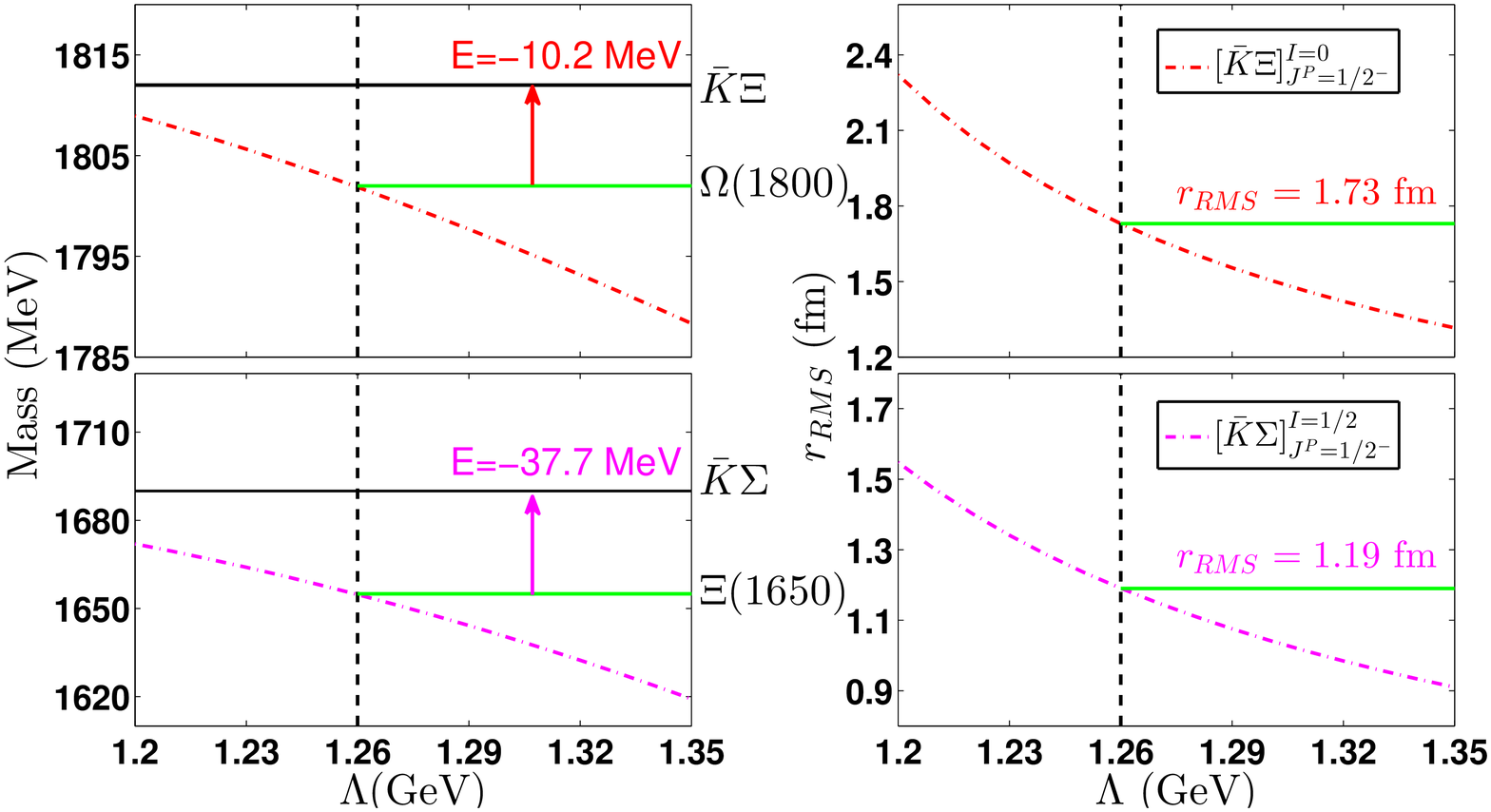}\\
\caption{The $\Lambda$ dependence of the bound state masses and the root-mean-square ($RMS$) radius $r_{RMS}$ for the $\bar{K}\Sigma$ and the $\bar{K}\Xi$ systems with quantum numbers $I(J^P)=1/2(1/2^-)$ and $I(J^P)=0(1/2^-)$, respectively. $E$ is the binding energy. The vertical dotted line and the horizontal solid lines correspond to the cutoff $\Lambda$ value and the thresholds of meson-baryon systems, respectively.}\label{threshold2}
\end{figure}

%As shown in Fig. \ref{threshold}, when cutoff is taken as $\Lambda=1.26$ GeV, we can obtain a good $\bar{K}N [0(1/2^-)]$ molecular state, with a binding energy $E=-30.9$ MeV and root-mean-square (RMS) radius $r_{RMS}=1.31$ fm. This molecular state corresponds to the $\Lambda(1405)$. If we take the same cutoff value, the $\bar{K}\Lambda$ state with $I(J^P)=1/2(1/2^-)$ can be a loose molecular candidate, its binding energy is $E=-2.9$ MeV, and the RMS radius is 3.52 fm. It is obvious that the mass of $\Xi(1620)$ observed by Belle Collaboration \cite{Sumihama:2018moz} is well produced. To summary, the newly observed $\Xi(1620)$ can be explained as a $\bar{K}\Lambda$ molecular state with $I(J^P)=1/2(1/2^-)$.
Taking the same cutoff $\Lambda=1.26$ GeV, we further study the $\bar{K}\Sigma$ and the $\bar{K}\Xi$ systems. As presented in Fig. \ref{threshold2}, the doubly strange $\bar{K}\Sigma[1/2(1/2)^-]$ state and the triply strange $\bar{K}\Xi[0(1/2)^-]$ state have the binding energies $-37.7$ MeV and $-10.2$ MeV, respectively. Here, in this work, these two predicted states are labeled as $\Xi(1650)$ and $\Omega(1800)$, respectively. We expect that further experiment can confirm our prediction to the existence of $\Xi(1650)$ and $\Omega(1800)$.
\renewcommand\tabcolsep{0.45cm}
\renewcommand{\arraystretch}{1.5}
\begin{table}[!htbp]
\caption{The contributions of the recoil effect when forming the $\bar{K}N[0(1/2)^-]$, the $\bar{K}\Lambda[1/2(1/2)^-]$, the $\bar{K}\Sigma[1/2(1/2)^-]$, and the $\bar{K}\Xi[0(1/2)^-]$ molecular states. The cutoff is fixed at $\Lambda=1.26$ GeV. Here, $E_{nor}$ and $E$ denotes the binding energy without and with including the recoil correction terms. All the binding energy are in units of MeV.}
\label{recoil}
\centering
\begin{tabular}{cccccc}
\toprule[1.5pt]
&$E_{nor}$&$E$\\
\hline
$\bar{K}N[0(1/2)^-]$&-20.4&-30.9\\
$\bar{K}\Lambda[1/2(1/2)^-]$&-3.3&-2.9\\
$\bar{K}\Sigma[1/2(1/2)^-]$&-27.4&-37.7\\
$\bar{K}\Xi[0(1/2)^-]$&-9.1&-10.2\\
\bottomrule[1.5pt]
\end{tabular}
\end{table}

Since we study the interactions of the systems composed of a light meson and a baryon, the recoil corrections may have considerable contributions when forming the molecular bound states. In Table \ref{recoil}, we present the binding energy with and without considering the recoil corrections for the $\bar{K}N[0(1/2)^-]$, the $\bar{K}\Lambda[1/2(1/2)^-]$, the $\bar{K}\Sigma[1/2(1/2)^-]$, and the $\bar{K}\Xi[0(1/2)^-]$ systems. The cutoff is fixed at $\Lambda=1.26$ GeV. %We can check the recoil correction contributions from different exchanged mesons, separately. In Eq .(\ref{VKN}), for the isospin singlet $\bar{K}N$ molecular state with quantum number $J^P=\frac{1}{2}^-$, the recoil term provides a repulsive force from $\sigma$ exchange potential, while the recoil corrections from $\rho$ and $\omega$ exchange potentials are both attractive. Thus, we finally obtain an attractive contribution with $E-E_{nor}=-10.5$ MeV from the recoil corrections of $\sigma$, $\rho$, and $\omega$ exchange potentials. This recoil effect plays an important role in forming the $\bar{K}N[0(1/2)^-]$ molecular state. Additionally, $\phi$ exchange potential will provide a repulsive recoil correction to the binding energy in Eq .(\ref{VKL}). Thus, by performing similar analyses, for the $\bar{K}\Lambda[1/2(1/2)^-]$, the $\bar{K}\Sigma[1/2(1/2)^-]$, and the $\bar{K}\Xi[0(1/2)^-]$ systems, all the numerical results of $E_{nor}$ in Table \ref{recoil} can also be well understood.}

Meanwhile, we show the binding energies and $RMS$ radii for the isospin partners of the $\bar{K}\mathcal{B}$ systems in Table \ref{result1}. The binding energy without including recoil corrections $E_{nor}$ are also presented. From Table \ref{result1}, we find that the recoil corrections still have considerable contributions in forming the $\bar{K}\Sigma[3/2(1/2)^-]$ and the $\bar{K}\Xi[1(1/2)^-]$ molecular states. In addition, for the $\bar{K}\Sigma$ state with $I(J^P)=3/2(1/2)^-$ and the $\bar{K}\Xi$ state with $I(J^P)=1(1/2)^-$, the binding energies are around several MeV and their $RMS$ radii are around several fm. When the cutoff $\Lambda$ is around $1.6$ GeV, we find that $E(\bar{K}\Sigma[1/2(1/2)^-])<E(\bar{K}\Sigma[3/2(1/2)^-])$ and $E(\bar{K}\Xi[0(1/2)^-])<E(\bar{K}\Xi[1(1/2)^-])$, which are reflected by that the interaction of $\bar{K}\Sigma[1/2(1/2)^-](\bar{K}\Xi[0(1/2)^-])$ is much stronger attractive than that of $\bar{K}\Sigma[3/2(1/2)^-](\bar{K}\Xi[1(1/2)^-])$.

\renewcommand\tabcolsep{0.27cm}
\renewcommand{\arraystretch}{1.5}
\begin{table}[!htbp]
\caption{Bound state solutions for the $\bar{K}\mathcal{B}$ systems. $E_{nor}$ and $E$ denotes the binding energy without and with including the recoil correction terms. Here, the cutoff $\Lambda$, the binding energy, and the root-mean-square radius $r_{RMS}$ are in units of GeV, MeV, and fm, respectively.}
\label{result1}
\centering
\begin{tabular}{cccc|cccc}
\toprule[1.0pt]
\multicolumn{4}{c|}{$\bar{K}\Sigma [3/2(1/2^-)]$}     &\multicolumn{4}{c}{$\bar{K}\Xi [1(1/2^-)]$}\\\hline
$\Lambda$   &$E_{nor}$ &$E$  &$r_{RMS}$   &$\Lambda$   &$E_{nor}$ &$E$    &$r_{RMS}$\\
1.5&-6.3&-3.0&3.38      &1.6 &-5.8&-2.1 &3.84\\
1.6&-10.8&-5.4&2.53      &1.9 &-13.0&-4.5 &2.80\\
1.7&-16.1&-8.0&2.20      &2.2 &-19.8&-6.2 &2.40\\\bottomrule[1.5pt]
\end{tabular}
\end{table}

\subsection{The $\bar{K}^*\mathcal{B}$ systems}\label{KSB}

\renewcommand\tabcolsep{0.1cm}
\renewcommand{\arraystretch}{1.8}
\begin{table}[!htbp]
\caption{Bound solutions for $\bar{K}^{*}\mathcal{B}$ systems. $E_{nor}$ and $E$ denotes the binding energy without and with including the recoil correction terms. Here, the cutoff $\Lambda$, the binding energy and the root-mean-square radius $r_{RMS}$ are in units of GeV, MeV, and fm, respectively.}
\label{result2}
\centering
\begin{tabular}{ccccc|ccccc}
\toprule[1.5pt]
$I(J^P)$&$\Lambda$&$E_{nor}$&$E$&$r_{RMS}$&$I(J^P)$&$\Lambda$&$E_{nor}$&$E$&$r_{RMS}$\\
\hline
\multicolumn{10}{c}{$\bar{K}^*\Lambda$}\\
\hline
$\frac{1}{2}(\frac{1}{2}^-)$ &1.1&-6.0&-4.9&2.36&$\frac{1}{2}(\frac{3}{2}^-)$&1.1&-4.7&-3.8&2.67\\
&1.2&-20.2&-16.8&2.36&&1.2&-16.3&-13.6&1.58\\
&1.3&-39.7&-33.0&1.16&&1.3&-32.2&-27.0&1.21\\
\hline
\multicolumn{10}{c}{$\bar{K}^*\Sigma$}\\
\hline
$\frac{3}{2}(\frac{1}{2}^-)$ &1.4&-5.3&-3.5&2.67&$\frac{3}{2}(\frac{3}{2}^-)$&1.05&-8.6&-3.9&2.56\\
&1.5&-9.5&-6.2&2.18&&1.15&-22.3&-14.7&1.50\\
&1.6&-14.3&-9.2&1.87&&1.25&-39.5&-29.4&1.15\\
\cline{2-10}
$\frac{1}{2}(\frac{1}{2}^-)$&0.88&-2.2&-1.8&3.36&$\frac{1}{2}(\frac{3}{2}^-)$&1.05&-2.9&-2.9&2.81\\
&0.93&-14.6&-13.6&1.51&&1.15&-16.9&-17.3&1.47\\
&0.98&-36.3&-35.2&1.05&&1.25&-39.0&-40.8&1.09\\
\hline
\multicolumn{10}{c}{$\bar{K}^*\Xi$}\\
\hline
$1(\frac{1}{2}^-)$&1.05&-7.0&-3.1&2.64 &$1(\frac{3}{2}^-)$&1.15&-3.7&-2.9&2.89\\
&1.15&-18.7&-11.8&1.60&&1.25&-9.8&-7.4&1.96\\
&1.25&-30.8&-21.7&1.26&&1.35&-17.4&-12.4&1.60\\
\cline{2-10}
$0(\frac{1}{2}^-)$&1.05&-2.6&-2.6&2.81&$0(\frac{3}{2}^-)$&1.0&-3.5&-2.9&2.87\\
&1.15&-17.1&-16.8&1.41&&1.1&-18.8&-18.0&1.37\\
&1.25&-40.0&-38.8&1.02&&1.2&-41.0&-40.0&1.02\\
\bottomrule[1.5pt]
\end{tabular}
\end{table}
For the $\bar{K}^*\mathcal{B}$ systems, the $S-D$ wave mixing effect is also considered, and the pseudoscalar exchange process is allowed. We list the bound state solutions in Table \ref{result2}. Here, the obtained conclusions include
\begin{enumerate}
  \item The recoil corrections still have considerable contributions on bounding the $\bar{K}^*\mathcal{B}$ systems.
  \item The $\bar{K}^*\Lambda$ systems with $I(J^P)=1/2(1/2^-)$ , $1/2(3/2^-)$ can be good candidates of doubly strange molecular states.
  \item For the $\bar{K}^*\Sigma$ systems, the states with $I(J^P)=1/2(1/2^-)$, $1/2(3/2^-)$, and $3/2(3/2^-)$ are promising molecular candidates. And $\bar{K}^*\Sigma[3/2(1/2)^-]$ as a molecular state is also possible.
  \item Several possible triply strange molecular states can be predicted, i.e., the $\bar{K}^*\Xi$ states with $1(1/2^-,3/2^-)$ and $0(1/2^-,3/2^-)$.
\end{enumerate}

Further, we also check the results when only considering $S$-wave contribution in the potentials. And we find that the above conclusions keep the same, as the $D$-wave contribution is negligible compared with the $S$-wave contribution.

\section{Summary}\label{sec4}

Searching for exotic hadronic matter is an interesting research issue for hadron physics. Especially, with more and more observations of charmoniumlike $XYZ$ states and $P_c$ states in the past years, the candidates of hidden-charm tetraquark and pentaquark have been provided, which also stimulated extensive discussions of different hadronic configurations \cite{Chen:2016qju,Liu:2019zoy,Guo:2017jvc,Hosaka:2016pey,Ali:2017jda,Karliner:2017qhf,Esposito:2016noz,Lebed:2016hpi,Richard:2016eis,Olsen:2017bmm}. Among them, hadronic molecular state is very popular to apply to explain these novel phenomena. Recently, the LHCb's observation of three $P_c$ states again gave strong evidence of hadronic molecular states composed of an anticharmed meson and a charmed baryon.

Besides the heavy flavor sector, theorists and experimentalists also paid more attentions to the light flavor sector. For example, the $\Lambda(1405)$ as a $\bar{K}N$ molecule with $I(J^P)=0(1/2^-)$ have been proposed \cite{Klempt:2009pi,Klempt:2007cp}.  Recently, Belle reported the observation of $\Xi(1620)$ \cite{Sumihama:2018moz} in the $\Xi_c^+\to \Xi^-\pi^+\pi^+$ process. If comparing the properties of the $\Xi(1620)$ and the $\Lambda(1405)$, we may find their similarities, which inspires our interest to exam the possibility of the newly observed $\Xi(1620)$ as
the $\bar{K}\Lambda$ molecular state.

In this work, we perform a systematical study on the $\bar{K}^{(*)}\mathcal{B}$ interactions within the framework of the one-boson-exchange model, where $\mathcal{B}$ stands for the strange or doubly strange ground octet baryons. Here, the $S-D$ wave mixing effect, the spin-orbit potential, and the recoil correction are taken into account. The analysis of recoil effect is explicitly performed. From our numerical results, we conclude that the recoil correction will provide considerable contributions for light-light $\bar{K}^{(*)}\mathcal{B}$ systems. By reproducing the mass of $\Lambda(1405)$ under the $\bar{K}N[0(1/2)^-]$ molecular picture, the parameter $\Lambda=1.26$ GeV can be fixed, which is directly applied to obtain the corresponding bound state solution for the $\bar{K}\Lambda$ molecular state. Our result shows that the newly observed $\Xi(1620)$ as the $\bar{K}\Lambda$ molecular state with $I(J^P)=1/2(1/2^-)$ can be supported in our theoretical framework.

Testing the scenario of the $\bar{K}\Lambda$ molecular assignment to the $\Xi(1620)$ is the main task of this work. In addition, we also give more theoretical predictions, i.e., there may exist the $\bar{K}\Sigma$ molecule with $I(J^P)=1/2(1/2^-)$ and the $\bar{K}\Xi$ molecule with $I(J^P)=0(1/2^-)$, which are labeled as $\Xi(1650)$ and $\Omega(1800)$, respectively.
Besides, the $\bar{K}^*\mathcal{B}$ systems are also investigated and some possible molecules composed of $\bar{K}^*\Lambda$, $\bar{K}^*\Sigma$, and $\bar{K}^*\Xi$ are also predicted.

Experimental search for these predicted states will be an interesting research topic.
More theoretical efforts should be paid in the near future. With the running of Belle II at Super KEKB, we have reason to believe that more evidence of light flavor molecular states will be revealed, which will provide more abundant information of exotic hadronic matter. It will be an effective way to deepen our understanding to the nonperturbative behavior of QCD.

\section*{Acknowledgments}
This project is supported by the China National Funds for Distinguished Young Scientists under Grant No. 11825503 and the National Program for Support of Top-notch Young Professionals. This project is also supported by the National Natural Science Foundation of China under Grant No. 11705069. R.C. is also supported by the National Postdoctoral Program for Innovative Talent.


\begin{thebibliography}{99}

%1
%\cite{Sumihama:2018moz}
\bibitem{Sumihama:2018moz}
  M.~Sumihama {\it et al.} (Belle Collaboration),
  Observation of $\Xi(1620)^0$ and Evidence for $\Xi(1690)^0$ in $\Xi_c^+ \rightarrow \Xi^-\pi^+\pi^+$ Decays,
  \href{https://journals.aps.org/prl/abstract/10.1103/PhysRevLett.122.072501}{Phys.\ Rev.\ Lett.\  {\bf 122}, 072501 (2019)}.
  %doi:10.1103/PhysRevLett.122.072501
  %[arXiv:1810.06181 [hep-ex]].
  %%CITATION = doi:10.1103/PhysRevLett.122.072501;%%


%2
%\cite{Bellefon:1900zz}
\bibitem{Bellefon:1900zz}
  A.~d.~Bellefon, A.~Berthon, and P.~Billoir,
  Reactions $K^- p \rightarrow \Xi^- K^{0(+)} \pi^{+(0)}$ between 2210 and 2435 MeV c.m.s energy,
  \href{https://link.springer.com/article/10.1007%2FBF02729820}
  {Nuovo Cimento A {\bf 28}, 289 (1975)}.



%3
%\cite{Ross:1972bf}
\bibitem{Ross:1972bf}
  R.~T.~Ross, T.~Buran, J.~L.~Lloyd, J.~H.~Mulvey, and D.~Radojicic,
  $\Xi^- \pi^+$ resonance with mass 1606 MeV/c$^2$,
  \href{https://www.sciencedirect.com/science/article/abs/pii/0370269372900433?via%3Dihub}
  {Phys.\ Lett.\  {\bf 38B}, 177 (1972)}.
  %doi:10.1016/0370-2693(72)90043-3
  %%CITATION = doi:10.1016/0370-2693(72)90043-3;%%
  %19 citations counted in INSPIRE as of 04 Apr 2019


  %4
 %\cite{Briefel:1977bp}
\bibitem{Briefel:1977bp}
  E.~Briefel {\it et al.},
  Search for $\Xi^*$ production in $K^- p$ interactions at 2.87 GeV/c,
  \href{http://inspirehep.net/record/118843}{Phys.\ Rev.\ D {\bf 16}, 2706 (1977)}.
  %doi:10.1103/PhysRevD.16.2706
  %%CITATION = doi:10.1103/PhysRevD.16.2706;%%
  %27 citations counted in INSPIRE as of 04 Apr 2019

  %5
   %\cite{Capstick:1986bm}
\bibitem{Capstick:1986bm}
  S.~Capstick and N.~Isgur,
  Baryons in a relativized quark model with chromodynamics,
  \href{https://journals.aps.org/prd/abstract/10.1103/PhysRevD.34.2809}{Phys.\ Rev.\ D {\bf 34}, 2809 (1986)}.
  %[AIP Conf.\ Proc.\  {\bf 132}, 267 (1985)].
  %doi:10.1103/PhysRevD.34.2809, 10.1063/1.35361
  %%CITATION = doi:10.1103/PhysRevD.34.2809, 10.1063/1.35361;%%
  %1169 citations counted in INSPIRE as of 25 Mar 2019




%6%\cite{GellMann:1964nj}
\bibitem{GellMann:1964nj}
  M.~Gell-Mann,
  A schematic model of baryons and mesons,
  \href{https://www.sciencedirect.com/science/article/pii/S0031916364920013?via%3Dihub}{Phys.\ Lett.\  {\bf 8}, 214 (1964)}.
  %doi:10.1016/S0031-9163(64)92001-3
  %%CITATION = doi:10.1016/S0031-9163(64)92001-3;%%
  %2848 citations counted in INSPIRE as of 04 Jun 2019

%7%\cite{Zweig:1981pd}
\bibitem{Zweig:1981pd}
  G.~Zweig,
  An SU(3) model for strong interaction symmetry and its breaking. Version 1,
  CERN, Report No. CERN-TH-401, 1964.
  %CERN-TH-401.
  %%CITATION = CERN-TH-401;%%
  %466 citations counted in INSPIRE as of 04 Jun 2019

  %8%\cite{Chen:2016qju}
\bibitem{Chen:2016qju}
  H.~X.~Chen, W.~Chen, X.~Liu, and S.~L.~Zhu,
  The hidden-charm pentaquark and tetraquark states,
  \href{https://www.sciencedirect.com/science/article/pii/S037015731630103X?via%3Dihub}{Phys.\ Rep.\  {\bf 639}, 1 (2016)}.
  %doi:10.1016/j.physrep.2016.05.004
  %[arXiv:1601.02092 [hep-ph]].
  %%CITATION = doi:10.1016/j.physrep.2016.05.004;%%
  %410 citations counted in INSPIRE as of 10 Jun 2019

%9%\cite{Liu:2019zoy}
\bibitem{Liu:2019zoy}
  Y.~R.~Liu, H.~X.~Chen, W.~Chen, X.~Liu, and S.~L.~Zhu,
  Pentaquark and tetraquark states,
  \href{https://www.sciencedirect.com/science/article/pii/S0146641019300304?via%3Dihub}{https://doi.org/10.1016/j.ppnp.2019.04.003}.
  %arXiv:1903.11976 [hep-ph].
  %%CITATION = doi:10.1016/j.ppnp.2019.04.003;%%
  %20 citations counted in INSPIRE as of 10 Jun 2019

%10%\cite{Guo:2017jvc}
\bibitem{Guo:2017jvc}
  F.~K.~Guo, C.~Hanhart, U.~G.~Mei{\ss}ner, Q.~Wang, Q.~Zhao, and B.~S.~Zou,
  Hadronic molecules,
  \href{https://journals.aps.org/rmp/abstract/10.1103/RevModPhys.90.015004}{Rev.\ Mod.\ Phys.\  {\bf 90}, 015004 (2018)}.
  %doi:10.1103/RevModPhys.90.015004
  %[arXiv:1705.00141 [hep-ph]].
  %%CITATION = doi:10.1103/RevModPhys.90.015004;%%
  %239 citations counted in INSPIRE as of 10 Jun 2019


  %11%\cite{Aaij:2019vzc}
\bibitem{Aaij:2019vzc}
  R.~Aaij {\it et al.} (LHCb Collaboration),
  Observation of a Narrow Pentaquark State, $P_c(4312)^+$, and of Two-Peak Structure of the $P_c(4450)^+$,
  \href{https://journals.aps.org/prl/abstract/10.1103/PhysRevLett.122.222001}{Phys.\ Rev.\ Lett.\  {\bf 122}, 222001 (2019)}.
  %doi:10.1103/PhysRevLett.122.222001
  %[arXiv:1904.03947 [hep-ex]].
  %%CITATION = doi:10.1103/PhysRevLett.122.222001;%%
  %27 citations counted in INSPIRE as of 11 Jun 2019



%12%\cite{Wu:2010jy}
\bibitem{Wu:2010jy}
  J.~J.~Wu, R.~Molina, E.~Oset, and B.~S.~Zou,
  Prediction of Narrow $N^*$ and $\Lambda^*$ Resonances with Hidden Charm Above 4 GeV,
  \href{https://journals.aps.org/prl/abstract/10.1103/PhysRevLett.105.232001}{Phys.\ Rev.\ Lett.\  {\bf 105}, 232001 (2010)}.
  %doi:10.1103/PhysRevLett.105.232001
  %[arXiv:1007.0573 [nucl-th]].
  %%CITATION = doi:10.1103/PhysRevLett.105.232001;%%
  %214 citations counted in INSPIRE as of 12 Jun 2019

%13%\cite{Yang:2011wz}
\bibitem{Yang:2011wz}
  Z.~C.~Yang, Z.~F.~Sun, J.~He, X.~Liu, and S.~L.~Zhu,
  The possible hidden-charm molecular baryons composed of anti-charmed meson and charmed baryon,
  \href{https://iopscience.iop.org/article/10.1088/1674-1137/36/1/002}{Chin.\ Phys.\ C {\bf 36}, 6 (2012)}.
  %doi:10.1088/1674-1137/36/1/002, 10.1088/1674-1137/36/3/006
  %[arXiv:1105.2901 [hep-ph]].
  %%CITATION = doi:10.1088/1674-1137/36/1/002, 10.1088/1674-1137/36/3/006;%%
  %111 citations counted in INSPIRE as of 12 Jun 2019

%14%%\cite{Chen:2019asm}
\bibitem{Chen:2019asm}
  R.~Chen, Z.~F.~Sun, X.~Liu, and S.~L.~Zhu,
  Strong LHCb evidence supporting the existence of the hidden-charm molecular pentaquarks,
  \href{https://journals.aps.org/prd/abstract/10.1103/PhysRevD.100.011502}
  {Phys.\ Rev.\ D {\bf 100}, 011502 (2019)}.
  %doi:10.1103/PhysRevD.100.011502
  %[arXiv:1903.11013 [hep-ph]].
  %%CITATION = doi:10.1103/PhysRevD.100.011502;%%
  %42 citations counted in INSPIRE as of 20 Aug 2019

%15
  %\cite{Chen:2019bip}
\bibitem{Chen:2019bip}
  H.~X.~Chen, W.~Chen, and S.~L.~Zhu,
  Possible interpretations of the $P_c(4312)$, $P_c(4440)$, and $P_c(4457)$,
  \href{https://journals.aps.org/prd/abstract/10.1103/PhysRevD.100.051501}{Phys.\ Rev.\ D {\bf 100}, 051501 (2019)}.
  %doi:10.1103/PhysRevD.100.051501
  %[arXiv:1903.11001 [hep-ph]].
  %%CITATION = doi:10.1103/PhysRevD.100.051501;%%
  %46 citations counted in INSPIRE as of 11 Oct 2019

%16%\cite{Liu:2019tjn}
\bibitem{Liu:2019tjn}
  M.~Z.~Liu, Y.~W.~Pan, F.~Z.~Peng, M.~Sanchez Sanchez, L.~S.~Geng, A.~Hosaka, and M.~Pavon Valderrama,
  Emergence of a Complete Heavy-Quark Spin Symmetry Multiplet: Seven Molecular Pentaquarks in Light of the Latest LHCb Analysis,
  \href{https://journals.aps.org/prl/abstract/10.1103/PhysRevLett.122.242001}
  {Phys.\ Rev.\ Lett.\  {\bf 122}, 242001 (2019)}.
 %doi:10.1103/PhysRevLett.122.242001
  %[arXiv:1903.11560 [hep-ph]].
  %%CITATION = doi:10.1103/PhysRevLett.122.242001;%%
  %42 citations counted in INSPIRE as of 20 Aug 2019

%17%\cite{He:2019ify}
\bibitem{He:2019ify}
  J.~He,
  Study of $P_c(4457)$, $P_c(4440)$, and $P_c(4312)$ in a quasipotential Bethe-Salpeter equation approach,
  \href{https://link.springer.com/article/10.1140%2Fepjc%2Fs10052-019-6906-1}{Eur.\ Phys.\ J.\ C {\bf 79}, 393 (2019)}.
  %doi:10.1140/epjc/s10052-019-6906-1
  %[arXiv:1903.11872 [hep-ph]].
  %%CITATION = doi:10.1140/epjc/s10052-019-6906-1;%%
  %21 citations counted in INSPIRE as of 12 Jun 2019

%18
%\cite{Huang:2019jlf}
\bibitem{Huang:2019jlf}
  H.~Huang, J.~He, and J.~Ping,
  Looking for the hidden-charm pentaquark resonances in $J/\psi p$ scattering,
  \href{http://inspirehep.net/record/1727564}{arXiv:1904.00221}.
  %%CITATION = ARXIV:1904.00221;%%
  %15 citations counted in INSPIRE as of 12 Jun 2019

  %19
  %\cite{Weng:2019ynv}
\bibitem{Weng:2019ynv}
  X.~Z.~Weng, X.~L.~Chen, W.~Z.~Deng, and S.~L.~Zhu,
  Hidden-charm pentaquarks and $P_c$ states,
  \href{https://journals.aps.org/prd/abstract/10.1103/PhysRevD.100.016014}
  {Phys.\ Rev.\ D {\bf 100}, 016014 (2019)}.
  %doi:10.1103/PhysRevD.100.016014
  %[arXiv:1904.09891 [hep-ph]].
  %%CITATION = doi:10.1103/PhysRevD.100.016014;%%
  %15 citations counted in INSPIRE as of 20 Aug 2019

%20
%\cite{Klempt:2009pi}
\bibitem{Klempt:2009pi}
  E.~Klempt and J.~M.~Richard,
  Baryon spectroscopy,
  \href{https://journals.aps.org/rmp/abstract/10.1103/RevModPhys.82.1095}{Rev.\ Mod.\ Phys.\  {\bf 82}, 1095 (2010)}.
  %doi:10.1103/RevModPhys.82.1095
  %[arXiv:0901.2055 [hep-ph]].
  %%CITATION = doi:10.1103/RevModPhys.82.1095;%%
  %284 citations counted in INSPIRE as of 10 Apr 2019

%21
%\cite{Klempt:2007cp}
\bibitem{Klempt:2007cp}
  E.~Klempt and A.~Zaitsev,
  Glueballs, hybrids, multiquarks. Experimental facts versus QCD inspired concepts,
  \href{https://www.sciencedirect.com/science/article/pii/S0370157307003560?via%3Dihub}{Phys.\ Rep.\  {\bf 454}, 1 (2007)}.
  %doi:10.1016/j.physrep.2007.07.006
  %[arXiv:0708.4016 [hep-ph]].
  %%CITATION = doi:10.1016/j.physrep.2007.07.006;%%
  %637 citations counted in INSPIRE as of 11 Apr 2019


%22
%\cite{Ramos:2002xh}
\bibitem{Ramos:2002xh}
  A.~Ramos, E.~Oset, and C.~Bennhold,
  On the Spin, Parity and Nature of the $\Xi$(1620) Resonance,
  \href{https://journals.aps.org/prl/abstract/10.1103/PhysRevLett.89.252001}{Phys.\ Rev.\ Lett.\  {\bf 89}, 252001 (2002)}.
  %doi:10.1103/PhysRevLett.89.252001
  %[nucl-th/0204044].
  %%CITATION = doi:10.1103/PhysRevLett.89.252001;%%
  %84 citations counted in INSPIRE as of 05 Apr 2019

%23
 %\cite{Miyahara:2016yyh}
\bibitem{Miyahara:2016yyh}
  K.~Miyahara, T.~Hyodo, M.~Oka, J.~Nieves, and E.~Oset,
  Theoretical study of the $\Xi$(1620) and $\Xi$(1690) resonances in $\Xi_c\rightarrow\pi+MB$ decays,
  \href{https://journals.aps.org/prc/abstract/10.1103/PhysRevC.95.035212}{Phys.\ Rev.\ C {\bf 95}, 035212 (2017)}.
  %doi:10.1103/PhysRevC.95.035212
  %[arXiv:1609.00895 [nucl-th]].
  %%CITATION = doi:10.1103/PhysRevC.95.035212;%%
  %13 citations counted in INSPIRE as of 05 Apr 2019

%24
%\cite{Wang:2019krq}
\bibitem{Wang:2019krq}
  Z.~Y.~Wang, J.~J.~Qi, X.~H.~Guo, and J. Xu,
  Analyzing $\Xi(1620)$ in the molecule picture in the Bethe-Salpeter equation approach,
  \href{https://link.springer.com/article/10.1140%2Fepjc%2Fs10052-019-7135-3}{Eur.\ Phys.\ J.\ C {\bf 79}, 640 (2019)}.
  %doi:10.1140/epjc/s10052-019-7135-3
  %[arXiv:1901.04474 [hep-ph]].
  %%CITATION = doi:10.1140/epjc/s10052-019-7135-3;%%
  %2 citations counted in INSPIRE as of 11 Oct 2019





%25
%\cite{Lu:2016nlp}
\bibitem{Lu:2016nlp}
  P.~L.~L¨¹ and J.~He,
  Hadronic molecular states from the $K\bar{K}^{\ast}$ interaction,
  \href{https://link.springer.com/article/10.1140%2Fepja%2Fi2016-16359-7}
  {Eur.\ Phys.\ J.\ A {\bf 52}, 359 (2016)}.
  %doi:10.1140/epja/i2016-16359-7
  %[arXiv:1603.04168 [hep-ph]].
  %%CITATION = doi:10.1140/epja/i2016-16359-7;%%
  %7 citations counted in INSPIRE as of 05 Apr 2019

%26
%\cite{Garzon:2012np}
\bibitem{Garzon:2012np}
  E.~J.~Garzon and E.~Oset,
  Effects of pseudoscalar-baryon channels in the dynamically generated vector-baryon resonances,
  \href{https://link.springer.com/article/10.1140%2Fepja%2Fi2012-12005-x}
  {Eur.\ Phys.\ J.\ A {\bf 48}, 5 (2012)}.
  %doi:10.1140/epja/i2012-12005-x
  %[arXiv:1201.3756 [hep-ph]].
  %%CITATION = doi:10.1140/epja/i2012-12005-x;%%
  %54 citations counted in INSPIRE as of 05 Apr 2019

%27
%\cite{Ding:2004ty}
\bibitem{Ding:2004ty}
  Y.~B.~Ding, X.~Li, X.~Q.~Li, X.~Liu, H.~Shen, P.~N.~Shen, G.~L.~Wang, and X.~Q.~Zeng,
  Estimating mass of sigma meson and study on application of the linear sigma model,
  \href{https://iopscience.iop.org/article/10.1088/0954-3899/30/7/001}
  {J.\ Phys.\ G {\bf 30}, 841 (2004)}.
  %doi:10.1088/0954-3899/30/7/001
  %[hep-ph/0402109].
  %%CITATION = doi:10.1088/0954-3899/30/7/001;%%
  %10 citations counted in INSPIRE as of 16 Aug 2019

%28
%\cite{Wang:2019aoc}
\bibitem{Wang:2019aoc}
  F.~L.~Wang, R.~Chen, Z.~W.~Liu, and X.~Liu,
  Possible triple-charm molecular pentaquarks from $\Xi_{cc}D_1/\Xi_{cc}D_2^*$ interactions,
  \href{https://journals.aps.org/prd/abstract/10.1103/PhysRevD.99.054021}{Phys.\ Rev.\ D {\bf 99}, 054021 (2019)}.
  %doi:10.1103/PhysRevD.99.054021
  %[arXiv:1901.01542 [hep-ph]].
  %%CITATION = doi:10.1103/PhysRevD.99.054021;%%
  %1 citations counted in INSPIRE as of 05 Apr 2019

%29
%\cite{Chen:2017jjn}
\bibitem{Chen:2017jjn}
  R.~Chen, A.~Hosaka, and X.~Liu,
  Prediction of triple-charm molecular pentaquarks,
  \href{https://journals.aps.org/prd/abstract/10.1103/PhysRevD.96.114030}
  {Phys.\ Rev.\ D {\bf 96}, 114030 (2017)}.
  %doi:10.1103/PhysRevD.96.114030
  %[arXiv:1711.09579 [hep-ph]].
  %%CITATION = doi:10.1103/PhysRevD.96.114030;%%
  %11 citations counted in INSPIRE as of 20 Aug 2019

%30
%\cite{Chen:2018pzd}
\bibitem{Chen:2018pzd}
  R.~Chen, F.~L.~Wang, A.~Hosaka, and X.~Liu,
  Exotic triple-charm deuteronlike hexaquarks,
  \href{https://journals.aps.org/prd/abstract/10.1103/PhysRevD.97.114011}
  {Phys.\ Rev.\ D {\bf 97}, 114011 (2018)}.
  %doi:10.1103/PhysRevD.97.114011
  %[arXiv:1804.02961 [hep-ph]].
  %%CITATION = doi:10.1103/PhysRevD.97.114011;%%
  %7 citations counted in INSPIRE as of 20 Aug 2019

  %31
  %\cite{Chen:2017xat}
\bibitem{Chen:2017xat}
  R.~Chen, A.~Hosaka, and X.~Liu,
  Searching for possible $\Omega_c$-like molecular states from meson-baryon interaction,
  \href{https://journals.aps.org/prd/abstract/10.1103/PhysRevD.97.036016}
  {Phys.\ Rev.\ D {\bf 97}, 036016 (2018)}
  %doi:10.1103/PhysRevD.97.036016
  %[arXiv:1711.07650 [hep-ph]].
  %%CITATION = doi:10.1103/PhysRevD.97.036016;%%
  %13 citations counted in INSPIRE as of 20 Aug 2019
%32
%\cite{Chen:2016ryt}
\bibitem{Chen:2016ryt}
  R.~Chen, J.~He, and X.~Liu,
  Possible strange hidden-charm pentaquarks from $\Sigma_c^{(*)}\bar{D}_s^*$ and $\Xi^{(',*)}_c\bar{D}^*$ interactions,
  \href{https://iopscience.iop.org/article/10.1088/1674-1137/41/10/103105}
  {Chin.\ Phys.\ C {\bf 41}, 103105 (2017)}.
  %doi:10.1088/1674-1137/41/10/103105
  %[arXiv:1609.03235 [hep-ph]].
  %%CITATION = doi:10.1088/1674-1137/41/10/103105;%%
  %8 citations counted in INSPIRE as of 20 Aug 2019
%33
%\cite{Liu:2011xc}
\bibitem{Liu:2011xc}
  Y.~R.~Liu and M.~Oka,
  $\Lambda_c N$ bound states revisited,
  \href{https://journals.aps.org/prd/abstract/10.1103/PhysRevD.85.014015}{Phys.\ Rev.\ D {\bf 85}, 014015 (2012)}.
  %doi:10.1103/PhysRevD.85.014015
  %[arXiv:1103.4624 [hep-ph]].
  %%CITATION = doi:10.1103/PhysRevD.85.014015;%%
  %67 citations counted in INSPIRE as of 26 Aug 2019

  %34
  %\cite{Meng:2017fwb}
\bibitem{Meng:2017fwb}
  L.~Meng, N.~Li, and S.~L.~Zhu,
  Deuteron-like states composed of two doubly charmed baryons,
  \href{http://inspirehep.net/record/1589437/export/hlxu}{Phys.\ Rev.\ D {\bf 95}, 114019 (2017)}.
  %doi:10.1103/PhysRevD.95.114019
  %[arXiv:1704.01009 [hep-ph]].
  %%CITATION = doi:10.1103/PhysRevD.95.114019;%%
  %15 citations counted in INSPIRE as of 26 Aug 2019

  %35
\bibitem{Greiner}
W.~Greiner and J.~Reinhardt,
\textsl{Field Quantization} (Springer-Verlag, Berlin, Heidelberg,1996), P155.

  %36
%\cite{Zhao:2014gqa}
\bibitem{Zhao:2014gqa}
  L.~Zhao, L.~Ma, and S.~L.~Zhu,
  Spin-orbit force, recoil correction, and possible $B \bar{B}^{*}$ and $D \bar{D}^{*}$  molecular states,
  \href{https://journals.aps.org/prd/abstract/10.1103/PhysRevD.89.094026}{Phys.\ Rev.\ D {\bf 89}, 094026 (2014)}.
  %doi:10.1103/PhysRevD.89.094026
  %[arXiv:1403.4043 [hep-ph]].
  %%CITATION = doi:10.1103/PhysRevD.89.094026;%%
  %25 citations counted in INSPIRE as of 05 Apr 2019

    %37%\cite{Lebed:2016hpi}
\bibitem{Lebed:2016hpi}
  R.~F.~Lebed, R.~E.~Mitchell, and E.~S.~Swanson,
  Heavy-quark QCD exotica,
  \href{https://www.sciencedirect.com/science/article/pii/S0146641016300734?via%3Dihub}{Prog.\ Part.\ Nucl.\ Phys.\  {\bf 93}, 143 (2017)}.
  %doi:10.1016/j.ppnp.2016.11.003
  %[arXiv:1610.04528 [hep-ph]].
  %%CITATION = doi:10.1016/j.ppnp.2016.11.003;%%
  %168 citations counted in INSPIRE as of 10 Jun 2019

%38%\cite{Richard:2016eis}
\bibitem{Richard:2016eis}
  J.~M.~Richard,
  Exotic hadrons: Review and perspectives,
  \href{https://link.springer.com/article/10.1007%2Fs00601-016-1159-0}{Few Body Syst.\  {\bf 57}, 1185 (2016)}.
  %doi:10.1007/s00601-016-1159-0
  %[arXiv:1606.08593 [hep-ph]].
  %%CITATION = doi:10.1007/s00601-016-1159-0;%%
  %38 citations counted in INSPIRE as of 10 Jun 2019

  %39%\cite{Olsen:2017bmm}
\bibitem{Olsen:2017bmm}
  S.~L.~Olsen, T.~Skwarnicki, and D.~Zieminska,
  Nonstandard heavy mesons and baryons: Experimental evidence,
  \href{https://journals.aps.org/rmp/abstract/10.1103/RevModPhys.90.015003}{Rev.\ Mod.\ Phys.\  {\bf 90}, 015003 (2018)}.
  %doi:10.1103/RevModPhys.90.015003
  %[arXiv:1708.04012 [hep-ph]].
  %%CITATION = doi:10.1103/RevModPhys.90.015003;%%
  %140 citations counted in INSPIRE as of 10 Jun 2019

%40%\cite{Karliner:2017qhf}
\bibitem{Karliner:2017qhf}
  M.~Karliner, J.~L.~Rosner, and T.~Skwarnicki,
  Multiquark states,
  \href{https://www.annualreviews.org/doi/10.1146/annurev-nucl-101917-020902}{Annu.\ Rev.\ Nucl.\ Part.\ Sci.\  {\bf 68}, 17 (2018)}.
  %doi:10.1146/annurev-nucl-101917-020902
  %[arXiv:1711.10626 [hep-ph]].
  %%CITATION = doi:10.1146/annurev-nucl-101917-020902;%%
  %44 citations counted in INSPIRE as of 10 Jun 2019

  %41%\cite{Esposito:2016noz}
\bibitem{Esposito:2016noz}
  A.~Esposito, A.~Pilloni, and A.~D.~Polosa,
  Multiquark resonances,
  \href{https://www.sciencedirect.com/science/article/pii/S037015731630391X?via%3Dihub}{Phys.\ Rep.\  {\bf 668}, 1 (2017)}.
  %doi:10.1016/j.physrep.2016.11.002
  %[arXiv:1611.07920 [hep-ph]].
  %%CITATION = doi:10.1016/j.physrep.2016.11.002;%%
  %193 citations counted in INSPIRE as of 10 Jun 2019

%42
 %\cite{Ali:2017jda}
\bibitem{Ali:2017jda}
  A.~Ali, J.~S.~Lange, and S.~Stone,
  Exotics: Heavy pentaquarks and tetraquarks,
  \href{https://www.sciencedirect.com/science/article/pii/S0146641017300716?via%3Dihub}{Prog.\ Part.\ Nucl.\ Phys.\  {\bf 97}, 123 (2017)}.
  %doi:10.1016/j.ppnp.2017.08.003
  %[arXiv:1706.00610 [hep-ph]].
  %%CITATION = doi:10.1016/j.ppnp.2017.08.003;%%
  %136 citations counted in INSPIRE as of 10 Jun 2019



%43
%\cite{Hosaka:2016pey}
\bibitem{Hosaka:2016pey}
  A.~Hosaka, T.~Iijima, K.~Miyabayashi, Y.~Sakai, and S.~Yasui,
  Exotic hadrons with heavy flavors: X, Y, Z, and related states,
  \href{https://academic.oup.com/ptep/article/2016/6/062C01/2240707}{Prog. Theor. Exp. Phys. {\bf 2016}, 062C01 (2016)}.
  %doi:10.1093/ptep/ptw045
  %[arXiv:1603.09229 [hep-ph]].
  %%CITATION = doi:10.1093/ptep/ptw045;%%
  %68 citations counted in INSPIRE as of 10 Jun 2019

\end{thebibliography}
\end{document}